\documentstyle[12pt,a4,epsfig]{article}

\begin{document}

\topmargin 0pt
\oddsidemargin 5mm

\setcounter{page}{1}
%\begin{titlepage}
\vspace{2cm}
\begin{center}

{\bf CHARGE ASYMMETRY IN 1-1000 GEV ELECTROMAGNETIC SHOWERS
AND POSSIBILITY OF ITS MEASUREMENT}\\
\vspace{3mm}
{\large  R.O.Avakian, K.A.Ispirian$^{*)} $, R.K.Ispirian, V.G.Khachatryan, 
P.Sona $^{a)} $ and E.Uggerhoj $^{b)}$ }\\
\vspace{3mm}
{\em Yerevan Physics Institute, Brothers Alikhanian 2, Yerevan, 375036, 
Armenia}\\
\vspace{3mm}
{\em a) INFN and University of Firenze, Firenze, Italy}\\
\vspace{3mm}
{\em b) Institute for Storage Ring Facilities, University of Aarhus, Denmark} 
\end{center}

\vspace{3mm}
\centerline{\bf{Abstract}}
For the high energy electromagnetic 
showers the thickness dependence of a) the development 
of electron and positron components, b) the difference 
between the secondary electron and positron numbers, c) the charge
asymmetry of high energy electromagnetic showers, as well as d) the spectral 
distributions of the components at the shower maxima for various energies 
of primary particle energies,1 - 1000 GeV were investigated employing GEANT 
Monte
Carlo simulation package. Using these simulation results it is 
discussed the possibility of observation and study of the charge asymmetry 
with the help of a magnetic spectrometer which is important for the current 
and future experiments on the detection of radiowaves produced by high energy 
neutrinos.



\vspace{10mm}

\indent
\centerline{\bf{I.Introduction}}

\indent
At present all the elementary electromagnetic processes taking place when high 
energy electrons and photons pass through  the matter are well known. Therefore, 
the 
formulation of the correct theory for high energy electromagnetic showers (EMS) 
is possible in principle.However, due to mathematical difficulties the 
construction of the EMS theory 

\indent
--------------------------------------------------------

\indent
*) e-mail: karo@robert.yerphi.am

\indent
is realized in various approximations, and almost always the study of the EMS 
theory and comparison with the experimental data
are carried out with the help of Monte Carlo simulalations using the existing 
codes as EGS, GEANT [1] and others developed at SLAC, CERN and other 
laboratories.
In all the available EMS codes the "fate" of electrons and  
photons is followed down to certain lowest energies, cut energies, in order 
to escape very large program complication and computation times. Usually these 
cut energies are less than 1 MeV. In some 
cases as in the study of biological processes and some processes considered 
below 
when high density energy depositions, say, by $ \delta $- or Compton electrons 
are essential one should lower these threshold energies. Nevertheless, there is 
a satisfactory  agreement between the experimental observations and Monte Carlo 
simulation results on EMS, and one can use the the laters to study some 
processes which are not observed and studied yet.

\indent
In 1961 G.Askarian [2] predicted an excess of electrons over positrons 
in high energy EMS due to positron annihilation in flight, Compton and $ \delta
$- electrons and estimated the intensity of coherent Cherenkov and transition 
radiation radiowaves produced by this moving negative charge excess. The 
estimates carried out in [2] neglecting the contribution of Compton-, 
$\delta $- and not mentioned in  [2] photo-electrons show that the number of 
electrons in EMS can exceed the number of positrons by more than 10  $ \% $ at 
energies higher than hundreds MeV. The more accurate calculations (see [3] and 
references therein) of such electron surplus with the help Monte Carlo 
simulations confirm the predictions of [2]. In the latest calculations [3] 
it has been shown that the following processes give main contribution 
in the production of the EMS charge excess: Compton scattering 
(50-60 $ \% $) , Bhaba scattering (30-35 $\% $), positron annihilation in flight 
(5-20 $\%$), and their contribution depends weakly on the primary 
particle energy and relatively strongly on the component energy. Due to the 
low cross section the number of MeV photons is 
larger than the combined number of electrons and positrons in EMS, and Compton 
effect gives the largest part of the excess in MeV energy region.However, at 
present there is no direct experimental results on the EMS charge excess, 
while the existing indirect data (see below) are ambigeous and need correct 
interpretation.

\indent 
In the work [2] it has been also estimated and shown that the intensity of the 
coherent Cherenkov radiation radiowaves produced by this moving negative 
charge excess is sufficient to be used for detection of high energy EMS on 
the earth and moon. Following [2] it was suggested [4] to use this radiowave 
production for the undergroud detection of high energy neutrinos in the salt 
mines, while the mechanisms of radiowave production has been considered in 
details by the authors of the work [5] assuming various time and space 
distributions of the charge excess. After these and later theoretical and 
experimental investigations carried out in sixties devoted to EMS charge 
excess and radiowave production (see, [6]) many works have been published 
by the cosmic ray physicists because the method promised to be very convinient
for the very high energy neutrino astrophysics and neutrino 
oscillation problems.

\indent
Despite the achievements in this field after more than 35 years of theoretical 
and experimental investigations and many 
interesting projects under construction (see [7] and Proceedings of last 
International Cosmic Ray Conferences), the technique of detection of EMS 
with the help on radio antennas has not yet been proven and difficulties are 
anticipated [8]. In this connection it seems reasonable to study experimentally 
the various characteristics of the EMS charge excess and of the coherent 
Cherenkov as well as transition radiation produced by  the available 
high energy electron beams at various accelerators.

\indent
        When this paper was ready for submitting an electronic preprint 
[9] has appeared in which the authors in addition to the existing 
experimental studies devoted to the far infrared and submillimeter 
coherent Cherenkov and transition radiation have investigated the 
polarization, angular, coherence and other properties of the same 
radiations in GHz radio region using 15.2 MeV electron bunches. The 
authors conclude that it is necessary to carry out more accurate 
measurement for various applications.

\indent
        Taking into account the above said, the actuality of the problem 
and the available contemporary computational possibilities  in this work we 
study the processes connected with the charge excess at primary energies 0.5 - 
256 GeV with the help of the GEANT code package. It is shown the possibility 
of the observation and experimental study  of this processes 
at YerPhI, SPS, CERN, and FERMILAB. 
        
\vspace{10mm}
\centerline{\bf{II.Results of Simulations}}

\indent
We have chosen GEANT [1] to carry out the necessary Monte Carlo simulations
 on EMS for two reasons. First, long term practice indicates that GEANT handles
 in a proper way. Second, GEANT is disigned to to simulate the geometry of the
 experimental setup, which is essential fo our purpose.  The 
agreement between our calculations and published results, in particular, 
on the depth dependence at higher energies, witnesses the correctness of 
our calculations. Calculations have been performed for various kinetic energy 
cuts
for electrons and photons, $ T_{cut} = T_{cut}^{e} = E_{cut}^{\gamma} 
$ from $ T_{cut} = 50 keV $ up to $ T = 12 MeV $ and when the primary particles 
were photons (the results for electrons do not differ significantly from those 
for photons) with total number $ N_{ \gamma}$ and various primary photon 
energies $ E_{\gamma}$ from 0.5 GeV up to 1000 GeV. Each 
element of the calculation array with fixed $ E_{\gamma}$ and $ E_{cut}$ 
containes information on a) the dependence of the electron and positron
numbers 
$ N _{e^-,e^+} $ upon the depth t in radiation length units; b) the dependence 
of the excess $ \nu = N_{e^-} - N_{e^+}) $ on t; c) the dependence 
of the charge assymetry $ A = (N_{e^-} - N_{e^+})/(N_{e^-} + N_{e^+})$ on t and 
d) the energy spectrum of the electrons at the depth where the maximum of the
charge excess for the given parameters takes place, $ t = t_{max} $.
\indent
All the calculations presented in this work have been carried out for BGO 
because it is a diamagnetic insulator, has a small radiation length unit 
(useful properties for radiowave detection), has sufficient scintillation 
yield which can be useful in 
some cases and is available.  Fig.1 shows the information of one array element 
when $ E_{\gamma} =128 $ GeV and $ T_{cut} =0.4 $ MeV. As it is seen from
Fig.1 a 
the showering behaviors for electrons and positrons are similar, but they differ 
significantly in magnitudes. The behavior of the excess t-dependence (see 
Fig.1b) 
reminds the usual behavior of shower curves with tails of the form $ exp(- 
\alpha t) $ and its more intense part around the maxima can be approximated 
roughly by the symmetric function $ \sim exp(-\omega_0 \tau^2)$ where $\tau 
= t-t_{max} $ and $ \alpha $ and $ \omega_0 $ are constants 
as it is suggested in [5] to calculate the 
radiation intensity. As it follows from Fig.1c for the given $ E_{cut} =
0.4 $ MeV the assymetry exceeds the value given in [2,3] because of the
contribution 
of low energy electrons produced due to Compton effect. However, as it 
will be shown below for energies of the electron component higher then few 
MeV the asymmetry becomes less than it is predicted in [2,3]. The results given 
in Fig.1d show that indeed one can measure the assymetry, and such measurement 
is easier for lower energies of electrons and positrons. As it is seen 
from Fig.2 the charge asymmetry virtually disappeares above 20 MeV. 

\indent
Using many such simulation results as ones presented in Fig.1 one can reveal 
the characteristic properties of the EMS charge asymmetry necessary for the 
future employment. In Fig.3 a and b it is given the dependence of the charge 
excess on $ E_{\gamma} $ (for fixed $ E_{cut} $) and $ E_{cut} $ (for fixed 
$ E_{\gamma} $), respectively. $ \nu $ increases almost linearly with 
the increase of $ E_{\gamma} $ and decreases with the decrease of $ E_{cut} $.
In Fig.4 a and b it is given given the dependence of the charge asymmetry on
$ E_{\gamma} $ (for fixed $ E_{cut} $) and $ E_{cut} $ (for fixed 
$ E_{\gamma} $), respectively. It is seen that $ A $ almost does not depend 
on $ E_{\gamma} $, decreases with the decrease of $ E_{cut} $ and almost 
vanishes when $ E_{cut} > 10 $ MeV. Therefore, it is advantageous to study 
the charge asymmetry at possible higher primary particle energies and lower 
cut energies. 

\indent
As expected the calculations show that at the shower maxima the contributions 
from various processes resulting in charge excess depend on the component 
energy, and in the energy region below few MeV where the number of the 
electrons and the charge asymmetry are larger the proportion of the 
contribution fron various processes coincides with that given in [3] for 
higher energies, $ E \geq 1 TeV $.

\vspace{10mm}

\indent
\centerline{\bf{III. Asymmetry Meassurement Using Magnetic Spectrometers}}

\indent
Various characteristics of EMS have been investigated experimentally for a wide 
energy region of electrons and photons from 50 MeV up to few TeV with 
accelerator and cosmic ray particles using various methods and detectors. 
The authors of the work [10] used streamer chambers. In all these works 
the measurements have been carried out for secondary particle energies not less  
than 1 MeV. This is not because the corresponding Monte Carlo calculations of 
that time were available for $ E_{cut} > 1 $ MeV, but because the applied 
methods did not allow to decrease further the energies of the detected 
secondary particles because of larger energy measurement errors due to 
multiple scattering.

\indent
The use of streamer chambers in magnetic field  with insulator layers (BGO) 
in which the EMS are developed seems more suitable for the EMS charge excess 
investigations because they give the possibility to carry out the measurements 
at various depths simultaneously. With low Z gas filling and magnetic fields 
B = 0.03 T the expected accuracy for the energy measurements are about 16, 
14, 7 and less than 5 $ \% $ for electron energies 0.075, 0.15, 0.3 and 1.0 
MeV, respectively. The use of streamer as well as time projection chambers is 
connected with technical difficulties, and it will be much easier to perform 
such studies with the help of low energy magnetic pair spectrometers. 

\indent
It will be convinient to carry out such an experiment with the  arrangement 
NA59, SPS, CERN (see Fig.6 of [11]) proposed for other purpose and which 
will 
be ready in spring 1999. Since as it has been mentioned above the charge excess 
characteristics do not depend whether the primary particles are electron or 
photons, the 150-180 GeV electrons of the H2 beamline or the gamma quanta 
produced by these electrons must be 
focused (The beam angular and energy spread are not important, while its cross 
section radius must be decreased to 1-2 cm, since the Mollier radius for BGO is 
2.4 
cm) on 5-15 cm thick BGO slabs replacing the berillium target in the experiment 
NA59, and the pair spectrometer magnet with B.l = 0.52 Tm (l is the length 
of the magnetic field) must be replaced with a weak magnet 
with B.dl = 0.0033 Tm. Such weak fields and few meter distance provide the 
deflection of particles with energies higher than few MeV under angles greater 
than the angles under which the particles leave the BGO radiator and detect 
them at transversal distances larger than few Moller radius. The higher 
energy components 
do not touch the sensitive parts of the detectors downstream the magnet. The 
energy of the photons in the region 96-144 GeV is determined by the tagging 
system.The energy of the negative and positive shower particles coming out from 
BGO is measured with the help of two or three drift chambers. Since there are 
no polarization measurements the charge excess measurements at some depth 
will be much easier than the shower measurements on the arrangement NA43
[12] using polarized photon beams.
Again the multiple scattering in various thin windows alowes to determine 
the charge for secondary particle energies higher than few MeV. Since the 
primary electron beam intensity is $ 4.10^5 min^{-1}$ and the expected number 
of photons with energies 96-144 GeV will be only less by one order, the 
measurement time at one depth estimated with the above given curves is 
about 1 and 10 hours in the cases of primary electrons and photons, 
respectively .

\indent
Though the number of shower components is much lower at GeV energies (see Fig.
2a), nevertheless, EMS charge excess measurements are also possible at such 
energies using the 
ejected electron and photon beams, say, at Yerevan Synchrotron because of 
their higher intensity, about $ 10^9 $ electrons or photons per second.
\vspace{10mm}

\indent
\centerline{\bf{IV.Discussion}}
In this work it is reported the results of the Monte Carlo simulations 
on the negative charge excess and differential energy spectra of secondary 
particles in EMS taking into account all the processes in the primary particle 
energy interval 1-256 GeV.
For the energy region of the shower components below few MeV (above ten MeVs) 
the presented results predict much larger (smaller) excess than the estimates 
[2,3]. Nevertheless, as it has been shown in this work this 
charge excess can be measured with the low intensity, but high energy ($ 
\geq $  100 
GeV ) electron and photon beams at SPS, CERN and Fermilab or using the high 
intensity but relatively low energy ( $ \geq $ 1 GeV) beams at YerPhI. At 
present 
since the expected intensities of the atmospheric EMS coherent radiation of 
very high  energy particles in the radio diapason is higher, than the intensity 
of the EMS transition radiation in the clouds [13] and the sensitivity 
threshold of antennas, the formers are detected unambiguously in coincidence 
with other extended shower detectors. However many problems concerning the 
correct mechanisms of radio wave production, spectral and angular distribution 
etc. remain unsolved before wide application in very high energy neutrino 
astrophysics. The results of the excess measurements proposed in this work can 
shed light on the problems of EMS radiowave detection in dense (ice or salt) 
and air media.

\newpage

\centerline{{\bf{Figure Captions}}}

\indent
Fig.1. Characteristic charge excess dependences in BGO for primary photons 
when $ E_{\gamma} = 128 $ GeV and $ E_{cut} = 0.4 $ MeV. a) Shower curves 
separately for secondary electrons (solid histograms) and positrons (dashed 
histogram); b) Dependence of the excess $ \nu = N_{e^-} - N_{e^+} $ upon t (in 
radiation length units): c) Dependence of the asymmetry  $ A = (N_{e^-} -
N_{e^+})/(N_{e^-} + N_{e^+}) $ upon depth, and d) Differential spectrum of
the electrons at the maximum of the charge excess (100 events are simulated).

\indent
Fig.2. The dependence of the charge asymmetry on the energy of electrons 
and positrons at the shower maximum in the energy intervals a)1-10 MeV and b)10-
50 MeV.

\indent
Fig.3. The dependence of the charge excess on a) $ E_{cut} $ for the fixed
$ E_{\gamma} = 128 $ GeV and on b) $ E_{\gamma} $ for the fixed 
$ E_{cut} = 0.4 $ MeV. 

\indent
Fig.4. The dependence of the charge asymmetry on a)  $ E_{cut} $ for the
fixed  $ E_{\gamma} = 128 $ GeV and on b) $ E_{\gamma} $ for the fixed
$ E_{cut} = 0.4 $ MeV.

\begin{figure}
\epsfig{file=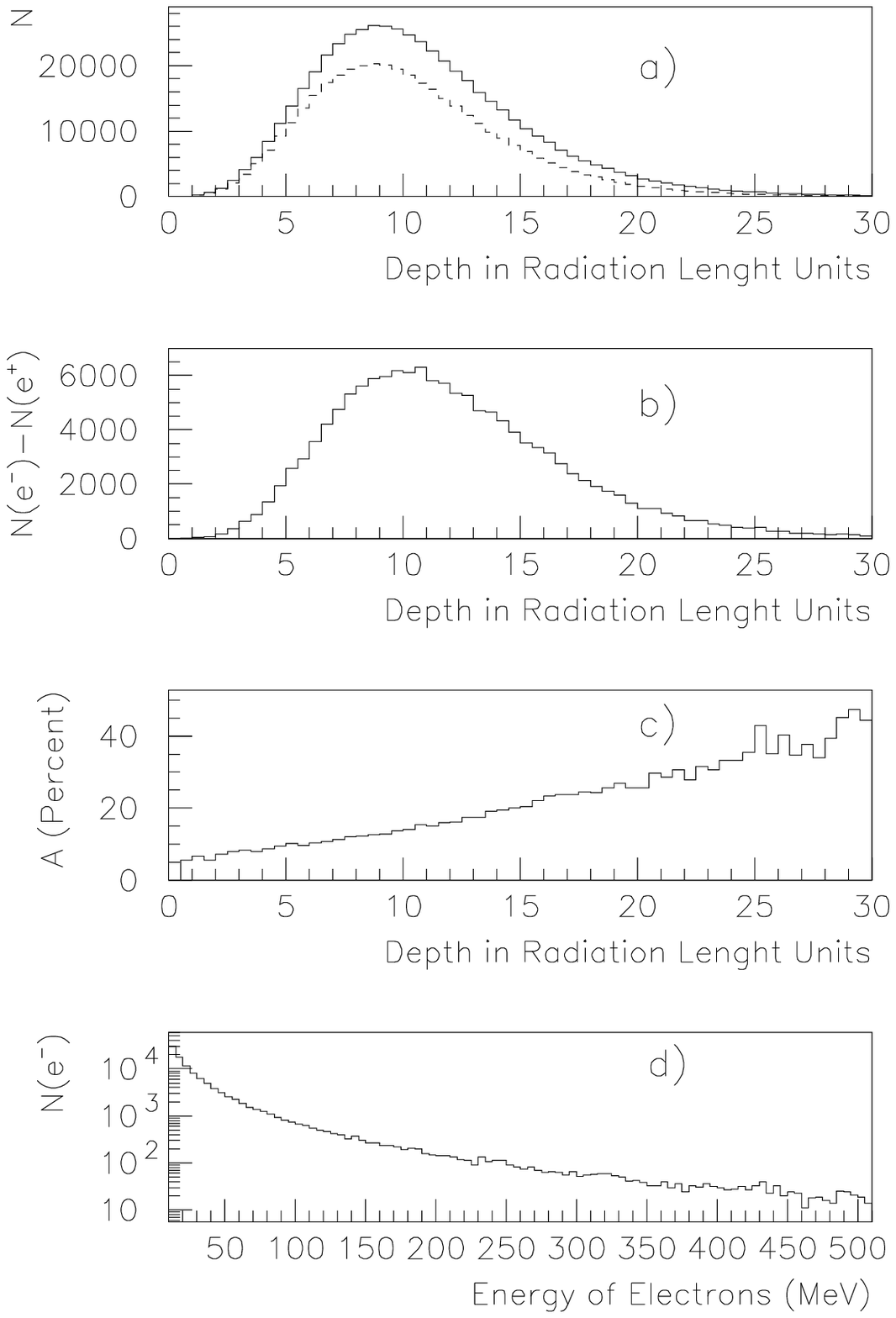,width=16cm,height=21cm}
\caption{}
\end{figure}

\begin{figure}
\epsfig{file=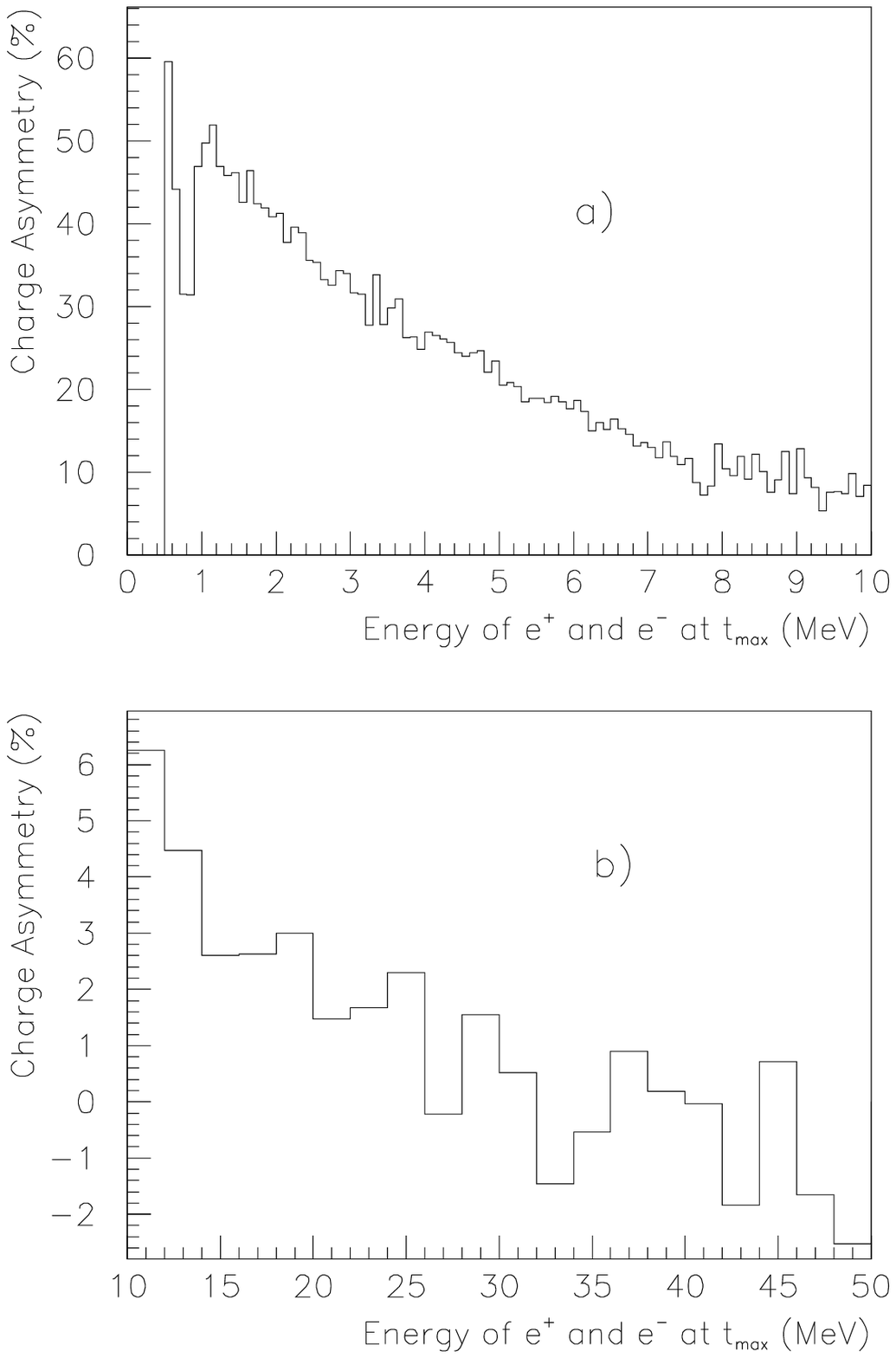,width=16cm,height=21cm}
\caption{}
\end{figure}

\begin{figure}
\epsfig{file=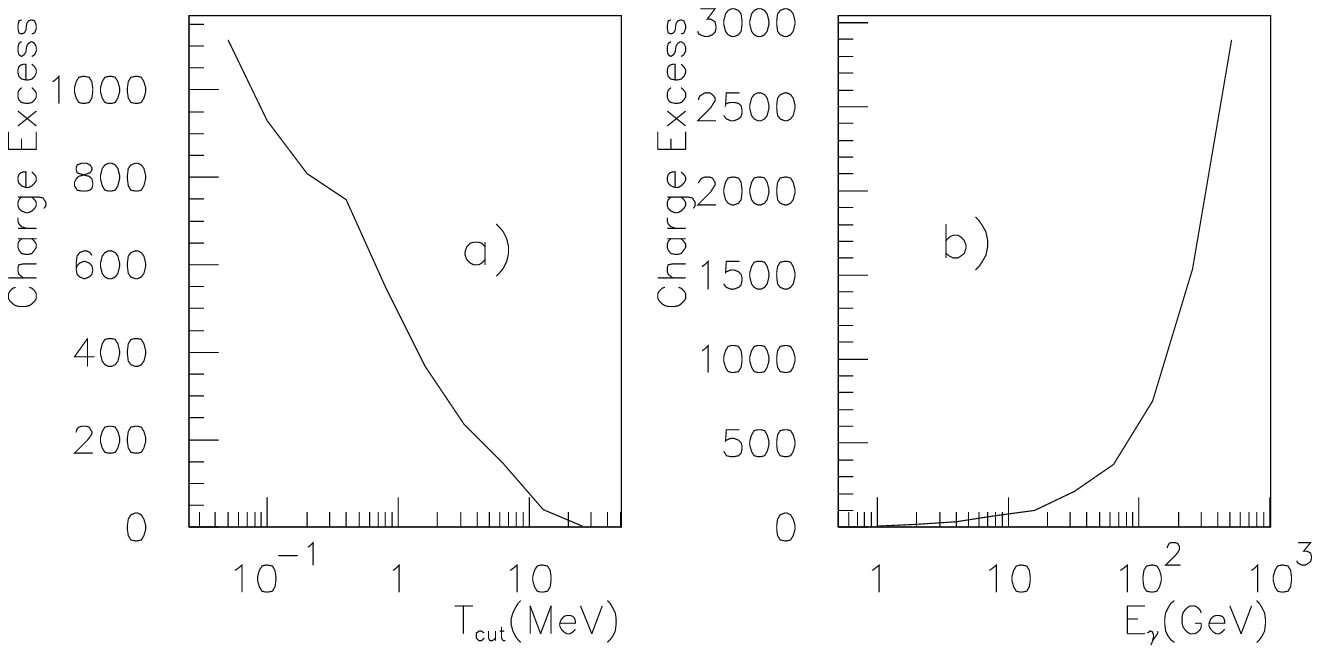,width=16cm,height=10cm}
\caption{}
\end{figure}

\begin{figure}
\epsfig{file=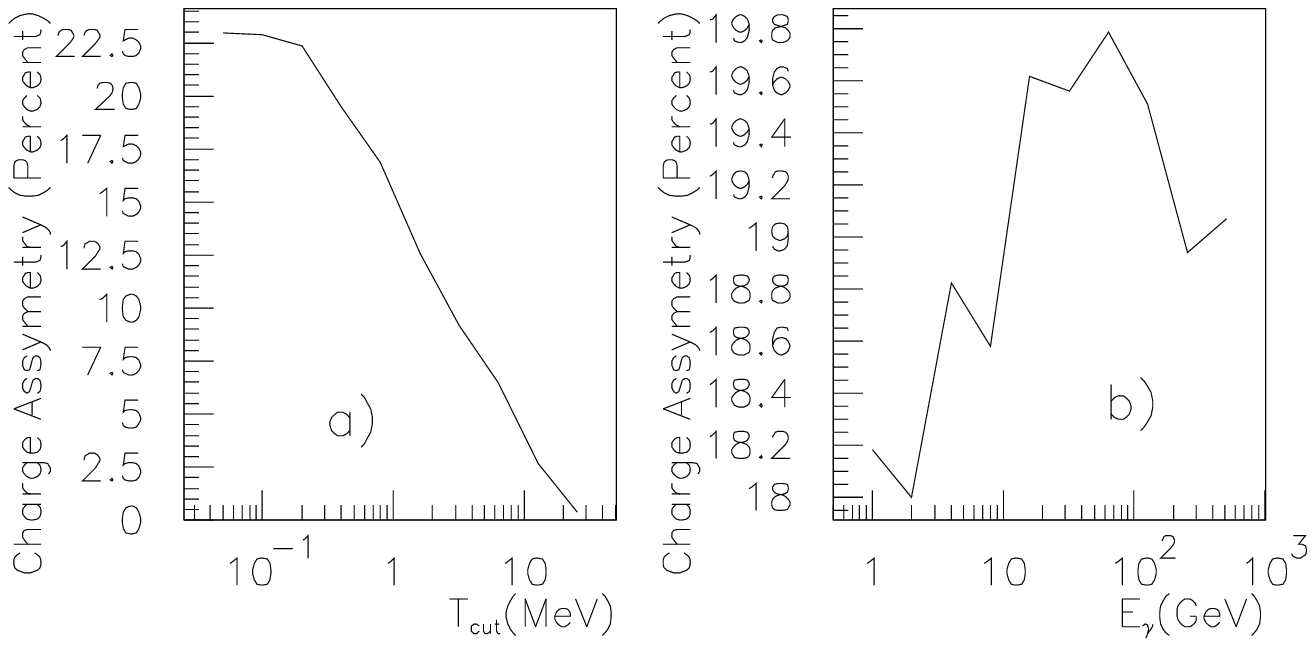,width=16cm,height=10cm}
\caption{}
\end{figure}

\end{document}